# Verification and Diagnostics Framework in ATLAS Trigger/DAQ

M.Barczyk, D.Burckhart-Chromek, M.Caprini[1], J.Da Silva Conceicao, M.Dobson, J.Flammer, R.Jones, A.Kazarov[2,3], S.Kolos[2], D.Liko, L.Lucio, L.Mapelli, I.Soloviev[2]
*CERN, Geneva, Switzerland*

R.Hart
*NIKHEF, Amsterdam, Netherlands*

A.Amorim, D.Klose, J.Lima, L.Pedro
*FCUL, Lisbon, Portugal*

H.Wolters
*UCP, Figueira da Foz, Portugal*

E.Badescu
*NIPNE, Bucharest, Romania*

I.Alexandrov, V.Kotov, M.Mineev
*JINR, Dubna, Russia*

Yu.Ryabov
*PNPI, Gatchina, Russia*

[1] *on leave from NIPNE*

[2] *on leave from PNPI*

[3] *paper editor*

Trigger and data acquisition (TDAQ) systems for modern HEP experiments are composed of thousands of hardware and software components depending on each other in a very complex manner. Typically, such systems are operated by non-expert shift operators, which are not aware of system functionality details. It is therefore necessary to help the operator to control the system and to minimize system down-time by providing knowledge-based facilities for automatic testing and verification of system components and also for error diagnostics and recovery.

For this purpose, a verification and diagnostic framework was developed in the scope of ATLAS TDAQ. The verification functionality of the framework allows developers to configure simple low-level tests for any component in a TDAQ configuration. The test can be configured as one or more processes running on different hosts. The framework organizes tests in sequences, using knowledge about components hierarchy and dependencies, and allowing the operator to verify the functionality of any subset of the system. The diagnostics functionality includes the possibility to analyze the test results and diagnose detected errors, e.g. by starting additional tests and understanding reasons of failures. A conclusion about system functionality, error diagnosis and recovery advice are presented to the operator in a GUI. The current implementation uses the CLIPS expert system shell for knowledge representation and reasoning.

## 1. INTRODUCTION

The ATLAS experiment [1] is one of four experiments at the Large Hadron Collider particle accelerator that is currently being built at CERN and is scheduled to start data taking in 2007. The ATLAS detector data rate and volume requires a very efficient Data Acquisition with a three-level trigger system [2].

The ATLAS Online Software (Online SW) is a subsystem of the ATLAS TDAQ project [2, chapter 5.3]. It encompasses the software to configure, control and monitor the TDAQ but excludes the processing and transportation of physics data. It must co-exist and co-operate with the other ATLAS sub-systems, in particular, interfaces are required to the data-flow, triggers, processor farms, event builder, detector read-out crate controllers and Detector Control System (DCS).

The Control subsystem of the Online SW includes software packages responsible for the distribution of run commands and synchronization between the systems, TDAQ initialization and shutdown, run supervision, error handling and diagnostics, system testing and verification, access management, process management and user interfaces.

The Diagnostics and Verification System (DVS) is one of the software packages of the Control subsystem. It is used for configuring and executing tests for TDAQ components, for detecting and diagnosing faults, and for advising recovery actions to the TDAQ operator.

## 2. DVS FUNCTIONALITY

The ATLAS TDAQ system has a very complex structure and behavior. Automation of system testing, error diagnosing and recovery are important issues for the control of such systems, because it helps to minimize experiment down-time.

DVS is a framework which allows TDAQ developers and experts to integrate tests and knowledge into it, so it can be later used by a non-experienced shift operator to verify the functionality of the TDAQ and diagnose problems. It is possible to have a number of tests defined for a single TDAQ component. Tests can be started on different hosts, sequentially or in parallel.

**TUGP005**

The main human users of DVS are the TDAQ Operator and TDAQ Expert. The functionality of DVS is also used by the TDAQ Supervisor application that helps the Operator to control the system.

DVS functionality is exploited by the users in the following cases (also presented in Figure 1):

- The TDAQ Expert implements and configures tests for TDAQ components and stores tests in a database. The Expert also stores the knowledge about testing sequences and components behavior in a knowledge base.
- During TDAQ initialization, TDAQ Supervisor application or TDAQ Expert launches a number of tests to ensure that hardware and software TDAQ components are correctly initialized.

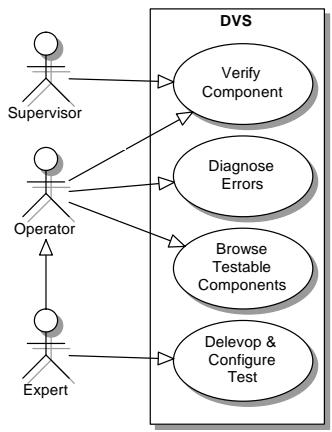

Figure 1. DVS users and functionality

- When an error is detected during the data taking, the TDAQ Operator can browse the TDAQ configuration in the DVS GUI and verify the status of a group of TDAQ components in order to detect problems. Using rules stored in the knowledge base and the test repository, DVS organizes and launches sequences of tests for selected components. Then it analyses test results, diagnoses errors and presents to the Operator a conclusion about the reason for the errors. Advice on how to repair failed components is also presented.

## 3. DVS DESIGN AND IMPLEMENTATION

### 3.1. Design approach

The main design ideas for DVS development were:

- to use *simple component tests*, developed by experts for TDAQ components
- to use *expert system* technology to store TDAQ developers *knowledge* in order to make it available for non-experienced shift operators
- develop a *framework* which allows to configure, store tests and store knowledge, which can be made available for later use by the operators
- develop end-user, *friendly GUI* application to be used by the operators

### 3.2. DVS package context

Figure 2 shows how DVS cooperates with users and other Online SW packages. The functionality of DVS can be used either by a human user (TDAQ Operator) via GUI or by other packages (TDAQ Supervisor) via API. To implement the required functionality, DVS reads the TDAQ configuration via the Configuration Databases service, launches tests via the Test Manager and uses the CLIPS package to implement the expert system.

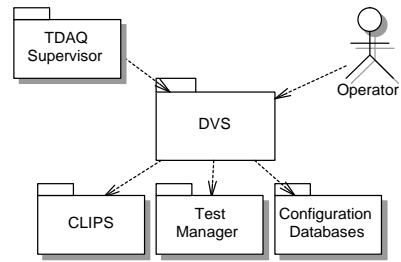

Figure 2. DVS package context

### 3.3. Implementation overview

The DVS internal architecture is presented in Figure 3. It is composed of a Test Repository Database, a Knowledge Base, an Expert System engine, C++ and Java libraries and a Graphical User Interface application.

The Test Repository and Expert System provide TDAQ developers (experts) with the possibility to:

- develop and configure tests for classes and objects in the TDAQ configuration, or redefine existing tests and store them in the test repository database
- develop the Knowledge Base, using the expert system language, to store specific knowledge about component functionality.

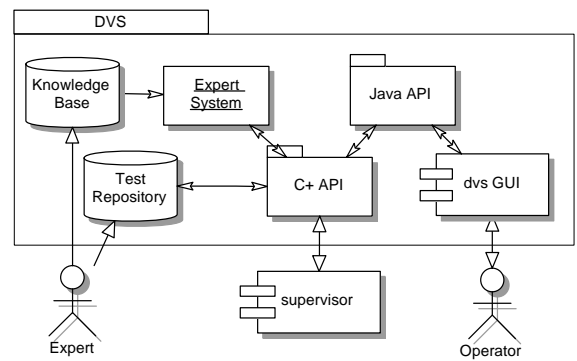

Figure 3. DVS package internal architecture

For end users (TDAQ Operator or other Online SW component like the Supervisor application), DVS provides the possibility to have a "testable" view on a TDAQ configuration, where a user can select a single component or a group of components and verify its status. This functionality is provided via GUI, or via the C++ and Java

APIs, so it can be used either directly by the operator or by another application.

### 3.4. What is a test

A test is a small application, which verifies status of a single software or hardware TDAQ component in a configuration and returns a test result (either Passed, Failed or Unresolved). Test should be as independent as possible, i.e. it should not rely on functionality of other TDAQ components. Typically a test is developed by a component expert. A test can be launched on any host used in the configuration. It is possible to have a number of tests defined for one TDAQ component, so they can be started on different hosts, synchronously one-by-one or in parallel.

Test processes are handled and executed with the help of the Test Manager and Process Manager [3, 4] - other Online SW components.

### 3.5. Test Repository

The Test Repository is a database which allows to describe different attributes of a test in the TDAQ Configuration Database [5].

The facilities provided by the TDAQ Configuration Databases are used to develop object schema, to store and to retrieve test objects from the database.

Each test in the repository is an instance of one of three classes defined in the Test Repository schema, presented in Figure 4: *Test*, *Test4Object* or *Test4Class*. These classes are used to describe test attributes and to associate the test with objects in the TDAQ Configuration database.

The base Test class describes basic test attributes:
- test implementation (as a link to a SW_object class from the TDAQ Configuration Database schema)
- test parameters
- test time-out
- host name where the test to be executed
- mode of test execution: synchronous or asynchronous (for the case where a number of tests are defined for a database object)
- order of tests execution (for the case where a number of tests are defined for a database object)

Test4Class and Test4Object classes, inherited from the Test class, are used to associate a test to objects in the TDAQ Configuration database. Instances of the Test4Object class are tests which verify the functionality of particular TDAQ components, whose database identifiers are stored in the "object_id" attribute of Test4Object. To define a test for all objects of a particular class, it is necessary to create an instance of Test4Class and fill its "class_name" attribute with the name of the class to be tested.

A C++ API (Test Data Access Library) is provided to access all the required configuration information.

The Test Repository and Test DAL are described in details in [6].

The Test Repository can contain tests for any TDAQ component described in a TDAQ configuration. Currently the Online SW test repository contains:
- tests for all TDAQ Online SW infrastructure applications
- a test for computer (remote access test)
- a test for VME module ("vme ping" test)
- a test for optical S-Link (source-destination test)

More tests are being implemented by TDAQ developers for their particular TDAQ subsystems and components. It is envisaged for the final TDAQ system to have a complete test repository, which covers all TDAQ components that can be tested.

### 3.6. Expert System

The core of DVS is an expert system engine, implemented in CLIPS ("C" language Integrated Production System) [7]. Its main features are:
- the expert system functionality is available via C API, so it can be integrated in C/C++ applications
- provides fully featured OO language
- uses "if-then" rules for knowledge representation
- free for non-commercial use
- available as source, easy to port to new platforms
- widely used, is known as the "de-facto" standard of forward-chaining rule-based (production) systems in the public domain

The DVS Knowledge Base (KB) is a number of text files with CLIPS object schema and rules. Currently it contains knowledge for testing and diagnosing application failures in the distributed TDAQ environment. There are rules to analyze results of testing, to start additional tests, to build diagnostics and advisory messages.

Users can extend KB by developing new classes and rules in CLIPS.

### 3.7. DVS GUI

The DVS GUI (shown in figures 5 and 6) presents the TDAQ configuration as a hierarchical tree of testable components (on the left side of the GUI). The user can select any component or a group of component and launch tests defined for these components. Test results, diagnosis of found errors (if any) and recovery advice are presented

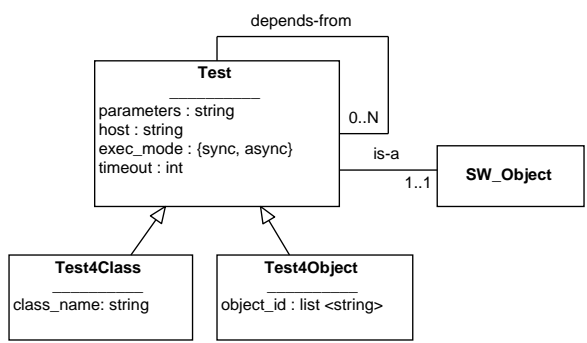

Figure 4. Test Repository schema

on the right side in separate panels for each tested component.

Other GUI features are:
- implemented in Java
- hypertext navigation over the output panels and the components tree
- log file browser for accessing log files produced by TDAQ applications running in a distributed environment
- help panel to read on-line HTML documentation for TDAQ components

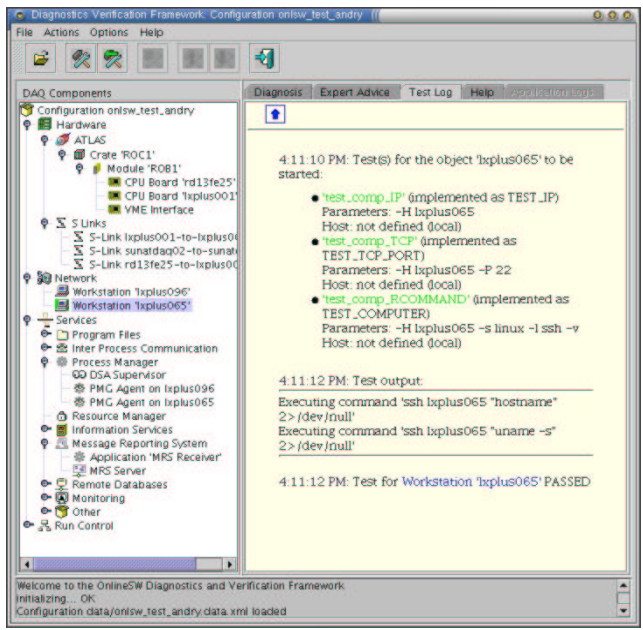

Figure 5. DVS GUI main window with test result

DVS GUI for the TDAQ configuration which is used for the current run can be launched from the main TDAQ Control "Integrated GUI" application [8].

### 3.8. Usage examples

In Figure 5 the screenshot of the DVS GUI is presented. It is an example of usage of the verification functionality of DVS. On the left side of the GUI one can see a tree of testable components in the loaded TDAQ configuration. One component (Workstation "lxplus075") was selected and tested. The test log is displayed in a hypertext panel on the right side. It shows a sequence of three independent tests launched for this workstation and the details of those tests, like parameters and host. Then the output of the tests and finally the test result ("PASSED") are displayed.

The screenshot in Figure 6 shows an example of failed testing for the "MRS Server" application (Online SW Message Reporting System server). To diagnose the failure, some additional tests were launched by DVS (according to rules in the KB) and the diagnosis of the failure was developed along with advices what to do in order to recover the failed component. The list of recovery actions to be applied by the operator to repair the problem with the "MRS Server" is presented in the "Expert Advice" panel on the right side of the GUI.

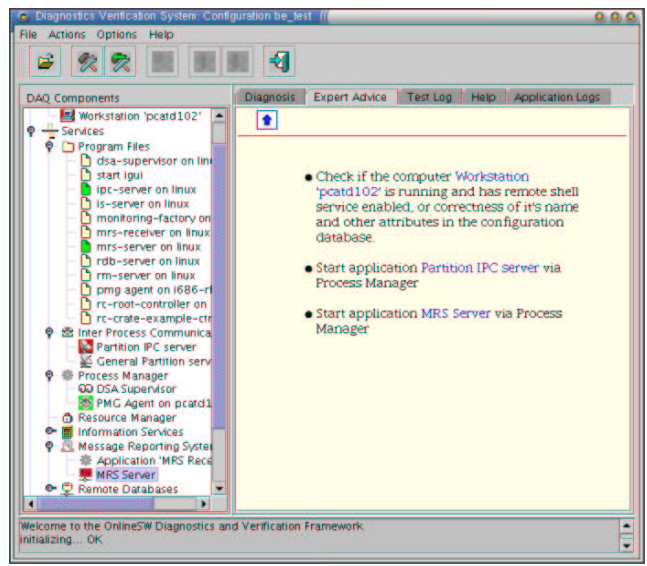

Figure 6. DVS GUI window with recovery advice

### 4. SUMMARY

The paper describes use cases, and the design and implementation details of the Diagnostics and Verification System of the ATLAS TDAQ system. DVS is a framework which is used for the configuration of tests for TDAQ components and for automation of their execution. Each TDAQ component in a configuration can be associated with a number of tests. Each test is a binary that can be launched on a computer in a distributed environment. All information about tests is stored in the Test Repository database. DVS is based on an expert system technique. Its knowledge base keeps TDAQ developers knowledge, useful for detecting and diagnosing faults and for advising the non-experienced TDAQ Operator of recovery actions.

More detailed information about DVS, including Users Guide can be found in [9].

### References

[1] ATLAS Collaboration, "Technical Proposal for a General-Purpose pp Experiment at the LHC collider at CERN", CERN/LHCC/94-43, 1994
[2] ATLAS Collaboration, "ATLAS High-Level Triggers, DAQ and DCS Technical Proposal", CERN/LHCC/2000-17, March 2000
[3] I.Alexandrov et al., "Process Management inside ATLAS DAQ", IEEE Transactions on Nuclear Science, Volume 29, Issue 5, Part 2, October 2002, pages 2459-2462
[4] R.Hart, "Implementation of Test Manager", ATLAS DAQ-1 Technical Note112, http://atddoc.cern.ch/Atlas/Notes/112/Note112-1.html

**TUGP005**

**TUGP005**